\documentclass[]{aastex}
\usepackage{emulateapj5_mod}
\newcommand{\myemail}{manojulu@tifr.res.in}
\shortauthors{Correlated radio:X-ray emission in microquasars}
\shorttitle{Choudhury et al.}
\defcitealias{cho02a}{Paper~1}
\submitted{Received 2002 November 05; Accepted 2003 April 22}

\begin{document}

\title{Correlated radio:X-ray emission in the hard states of Galactic microquasars}
\author{M. Choudhury\altaffilmark{1}, A. R. Rao, S. V. Vadawale}
\affil{Tata Institute of Fundamental Research, Mumbai-400005, India}
\author{A. K. Jain}
\affil{ISRO Satellite Centre, Bangalore-560017, India}
\altaffiltext{1}{For off prints contact M. Choudhury \myemail} 

\begin{abstract}
We present results of our study of correlated radio and X-ray emission in two black
hole candidates and Galactic microquasars GRS 1915+105 and Cyg X-1 in their
steady long term hard states, along with Cyg X-3 (using data obtained from {\it
Rossi X-Ray Timing Explorer} all-sky monitor [{\it RXTE}-ASM], {\it Compton Gamma Ray
Observatory} Burst and Transient Source Experiment [{\it CGRO}-BATSE], and Green Bank
Intereferometer [GBI]). We detect a pivotal behavior in the X-ray spectrum
of GRS 1915+105, correlated to the radio emission. Similar to the results obtained
for Cyg X-3, the flux of X-rays softer than the pivoting point correlates with
the radio emission, while the corresponding harder X-ray flux anti-correlates with
both the radio and the softer X-ray emission, in this state. We examine all the
previously reported correlations of X-ray properties with the radio emission in
Galactic microquasars and argue that these are consistent with a general picture
where a spectral pivoting is a common feature in these sources with the shape of
the spectrum determining the flux of radio emission, during the hard states. We
also detect a general monotonic increase in the radio emission of these sources with
the soft X-ray emission spanning about 5 orders of magnitude. We qualitatively
explain these findings with a Two Component Advective Flow model where the location
of a boundary layer between the thin disk and the Comptonizing region determines the
spectral shape and also the amount of outflow.
\end{abstract}


\section {Introduction}
The (quasi) simultaneous observations of X-ray binaries in the radio and X-ray bands
has led to the notion that the presence of radio jets is ubiquitous in sources with
black holes or low magnetic field ($\lesssim 10^9$ G) neutron stars as compact
objects \citep[see][]{fen01a, kul01}. Galactic X-ray binaries exhibiting radio jets
(outflow of matter in a collimated beam), with both physical as well as temporal
(variability) scale roughly at 6 orders of magnitude less than those of quasars, are
termed as {\it microquasars} \citep{mir99}. 

Although superluminally moving radio jets are detected in several microquasars
\citep{mir94,tin95,hje95}, which are invariably associated with huge radio
flares, only recently it has been realized that non-thermal radio emission is a
common feature during relatively quiet phases. \citet{fen01a,fen01b} made a detailed
calculation of the energetics during such quiet phases and argues that the
non-thermal emission forms a substantial part (5\%---50\%) of the energy budget.
Compact radio jets are indeed observed in the low-hard state of several microquasars
viz. Cyg X-1 \citep{sti01}, GRS 1915+105 \citep{dha00}, 1E 1740.7-2942
\citep{mir92}, and GRS 1758-258 \citep{rod92}. Further, the spectral analysis of the
radio emission from X-ray transient blackhole candidates, GS 2023+38, GRO J0422+32
and GS 1354-64 \citep{fen01a} and two persistent X-ray blackhole candidates GX 339-4
\citep{cor00} and XTEJ 1550-564 \citep{cor01} are interpreted to originate from
synchrotron emitting, compact conical jets. A comprehensive study of the association
between radio and X-ray emission during the quiet phases of these sources is
imperative for understanding the physical mechanism connecting the inflow and outflow
of matter, {\it a.k.a.} the disk-jet connection, in such systems.

The Galactic blackhole candidate GX 339-4 exhibits canonical X-ray states typical for
such binary systems. In its low-hard state the source shows very strong correlation
between radio and both soft and hard X-ray emissions \citep{cor03}, whereas in the
high-soft state the radio emission is quenched \citep{cor00}. For Cyg X-1, the
archetypical blackhole binary system, \citet{bro99} report that the radio flux is
loosely correlated with both soft X-ray (2 -- 12 keV, all-sky monitor [ASM]) and hard
X-ray (20 -- 100 keV, Burst and Transient Source Experiment [BATSE]) fluxes in the
low-hard state of the source, with the soft (ASM) and hard (BATSE) X-ray flux being
significantly correlated. The radio emission is suppressed in the high-soft state of
this source. The enigmatic blackhole candidate GRS 1915+105 shows several variability
classes, see \citet{bel00} for the various classifications, of which the class $\chi$
with steady low X-ray flux has the closest association to the canonical low-hard
state \citep{vad01a}. For this source, \citet{rau03} find no correlation between the
radio emission and both soft X-ray emission and wide-band (1 - 200 keV) flux, but
find an anti-correlation between hard X-ray (20 - 200 keV) flux and the radio
emission. They also find that the slope of the hard power law spectrum correlates
positively with the radio flux in this $\chi$ state. Thus, different types of radio
and X-ray correlations have been reported for these sources and a clear and
consistent picture of radio and X-ray flux correlation in the low-hard state for
these black hole candidates is yet to emerge.

Recently, we have carried out a systematic analysis of correlation between radio,
soft, and hard X-ray fluxes as well as a study of changes in wide band X-ray spectral behavior with radio flux in the steady quiescent X-ray state for Cyg X-3
\citep[henceforth Paper~1]{cho02a}, a suspected black hole candidate with strong
radio emission from a jet \citep{mio01, mart01}. The X-ray emission of Cyg X-3 quite
distinctively shows low and high states \citep{raj94,nak93}, which correspond to hard
and soft states \citep{cho02b}, distinguished by the shape of the X-ray spectra
characterized, chiefly but not totally, by the presence (or absence) of multicoloured
disk blackbody component and the power-law index (albeit with the individual model
components more complicated than the canonical X-ray states of classical black hole
candidates characterized, chiefly, by Cyg X-1, see, for eg. \citealt{tan95}). In
\citetalias{cho02a} we reported a strong and significant correlation between the
radio and the soft X-ray emission (2 -- 12 keV, [{\it RXTE}-ASM] in the hard state of
Cyg X-3, while the hard X-ray (20 -- 100 keV, {\it CGRO}-BATSE) anti-correlates with
both the soft X-ray and the radio, the last anti-correlation was first reported by
\citet{mcc99}. The wide-band X-ray spectral analysis of the pointed {\it RXTE}
observations in this hard state of the source at different radio flux levels show a
definite pivoting of the spectrum around 12 keV correlated to the radio emission
\citepalias[see Figure 3 and Table 2 of][]{cho02a}. As the interrelationship between
radio and X-ray flux reported in \citetalias{cho02a} has significant implications for
connectivity between accretion disk and radio jet, we have explored other black hole
candidates, for which similar radio and X-ray data are available, for radio and X-ray
flux correlation analysis.  

Here we present results of our analysis for the two persistent X-ray binaries,
microquasars and black hole candidates, Cyg X-1 and GRS 1915+105 for which
quasi-simultaneous radio and X-ray data are available from GBI, {\it RXTE}-ASM, and
{\it CGRO}-BATSE. We select data during periods when there are no radio flares and
the source is bright both in radio and hard X-rays. We examine radio X-ray
correlations in such hard states and show that these correlations are similar to
those found in the low-hard states of well studied sources like GX 339-4. We
complement our analysis of this state by giving a qualitative self consistent picture
of the steady X-ray non-flaring states of these sources, along with Cyg X-3 and GX
339-4, using the Two Component Advective Flow (TCAF) model of \citet{cha96}.

\section {Data and Analysis}
\citet{kul01} have compiled an extensive list of X-ray binaries, both neutron stars
and black hole candidates, for which simultaneous X-ray and radio observations have
been made. The Green Bank Interferometer, West Virginia, operated by NRAO, provides
data for a number of X-ray sources that were monitored during its several  years of
operation. Collating these sources with those monitored by {\it RXTE}-ASM and {\it
CGRO}-BATSE we found two X-ray binaries (black hole candidates) viz. Cyg X-1 and GRS
1915+105, along with Cyg X-3, which are persistent in radio, soft and hard X-ray
bands and for which (quasi) simultaneous data from the three observatories are
available.

\begin{figure}
\figurenum{1}
\epsscale{0.6}
\plotone{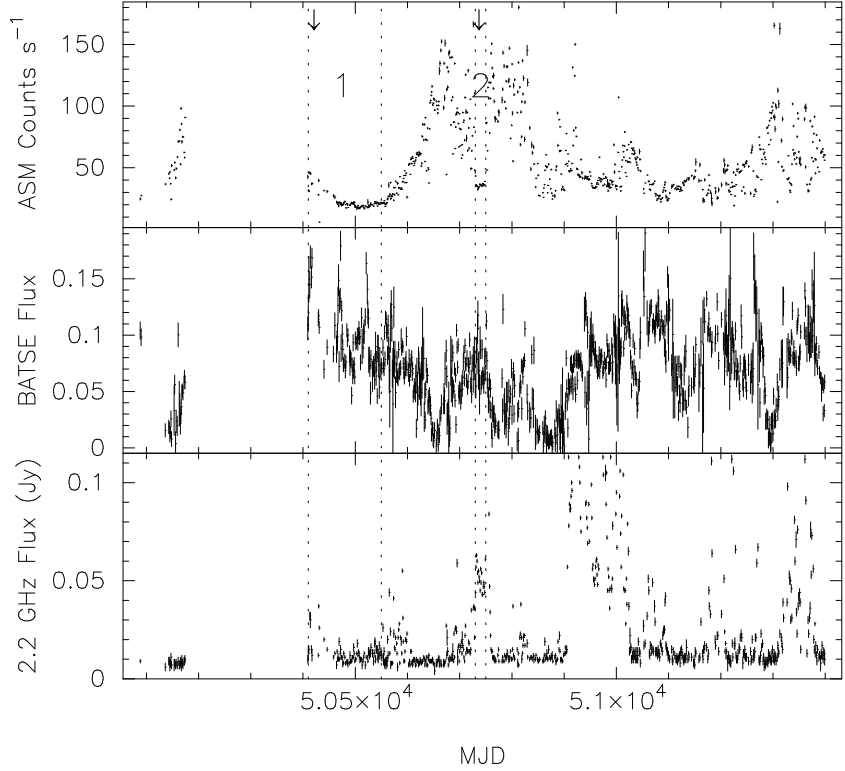}
\caption{The combined simultaneous light curve of GRS 1915+105 in the soft X-ray
(2 -- 12 keV, ASM, {\it top panel}), hard X-ray (20 -- 100 keV, BATSE, {\it middle panel}) and
the radio (2.2 GHz, GBI, {\it bottom panel}). The low-hard states, selected for the
present analysis (see text), are separated by vertical dashed lines and identified
with numbers. \label{fig1}}
\end{figure}

\begin{figure}
\figurenum{2}
\epsscale{0.6}
\plotone{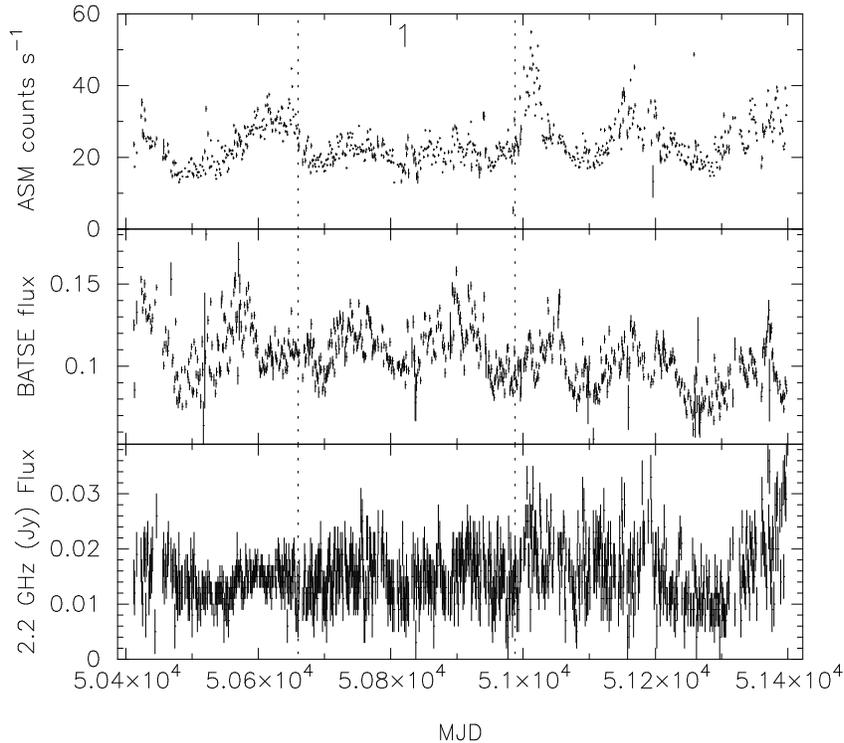}
\caption{The combined simultaneous light curve of Cyg X-1 in the soft X-ray
(2 -- 12 keV, ASM, {\it top panel}), hard X-ray (40 -- 140 keV, BATSE, {\it middle panel}) and
the radio (2.2 GHz, GBI, {\it bottom panel}). The low-hard state of the source, selected
for the present analysis (see text), is separated by vertical dashed lines and
identified with numbers. \label{fig2}}
\end{figure}

Figure \ref{fig1} and Figure \ref{fig2} give the daily averaged light curves of GRS
1915+105 and Cyg X-1, respectively, in the soft X-ray (ASM, {\it top panel}), hard
X-ray (BATSE, {\it middle panel}) and the radio (2.2 GHz, GBI, {\it bottom panel})
during the period when all the three instruments were simultaneously monitoring the
sources. Analogous to our approach for testing the correlation among the radio, soft
and hard X-ray emission from Cyg X-3 \citepalias{cho02a}, we have used the Spearman
Rank Correlation (SRC) test and adapted the method of partial rank correlation to
test the influence of the third parameter \citep{mac82}. The SRC coefficient gives
the strength of the correlation between two variables and the corresponding
D-parameter gives the confidence level, in terms of standard deviation, that the
derived correlation is independent of the influence of the third parameter. Here we
also give the results of the correlation analysis for data averaged over different
time intervals viz., 1, 5 and 10 days as well as correlation between X-ray hardness
ratio (ratio of observed BATSE count rate to that of ASM count rate) and radio (ASM
count rate is taken as the third parameter) for the duration in which the respective
sources were in a relatively long term steady quiescent hard state. Table \ref{tab1}
gives the complete result of the Spearman Rank Correlation (SRC) test for the
correlation among 1) the radio, soft X-ray and hard X-ray, and 2) the hardness ratio
in X-ray and radio (GBI) for GRS 1915+105 and Cyg X-1 along with  Cyg X-3, during the
steady quiescent hard state of X-ray emission, for fluxes in different bands averaged
over 1, 5 and 10 days.

\begin{figure}
\figurenum{3}
\epsscale{0.5}
\plotone{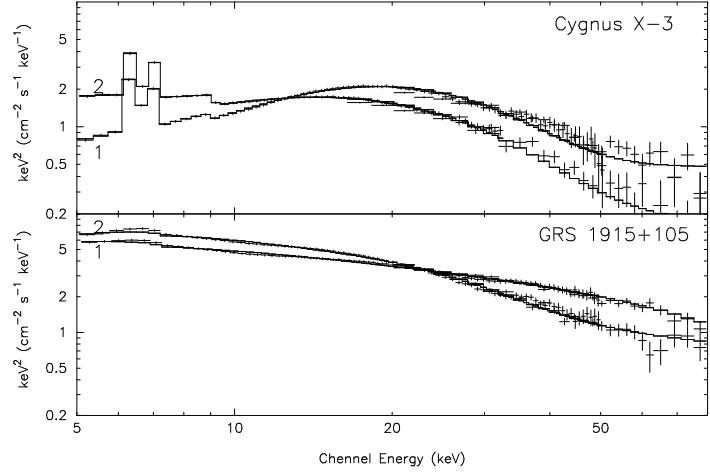}
\caption{{\it Upper Panel:} Unfolded spectra of Cyg X-3 in the low-hard state
(radio quiescent period) on two occasions (1: MJD 50954; 2: MJD 50717). {\it Lower
Panel:} The spectra of GRS 1915+105 in low-hard state on two occasions (1: MJD
50421; 2: MJD 50737). \label{fig3}}
\end{figure}

As in \citetalias{cho02a} for Cyg X-3, a detailed wide band spectral analysis of GRS
1915+105 has been done using the pointed mode observations by {\it RXTE} Proportional
Counter Array (PCA) and High Energy X-ray Timing Experiment (HEXTE) instruments
available in {\it RXTE} archive during the period of radio monitoring of this source
by GBI. For this purpose, two sets of X-ray data in the $\chi$ (hard) state of the
source and corresponding to two extreme values of the radio flux, within the of
precincts of this state, have been used for detailed spectral analysis following the
method described in \citetalias{cho02a}. These representative observations,
corresponding to the extreme behavior of the sources within the precincts of the
respective low and hard states, are marked as inverted arrows in Figure \ref{fig1}
and some features of these observations are given in Table \ref{tab2}. Two unfolded
spectra for GRS 1915+105 are overlaid on bottom panel of Figure \ref{fig3} with top
panel showing similar spectra for two extreme values of radio flux in the similar 
state for Cyg X-3 for comparison purpose \citepalias[from][]{cho02a}.

\subsection {Cyg X-3}
Cyg X-3 is one of the brightest radio sources associated with an X-ray binary. It
exhibits radio flares reaching up to 20 Jy in intensity, with a steady emission in
the range  of 40 -- 150 mJy during the quiescent state \citep{wal95}. Table
\ref{tab1} gives the result of the Spearman Rank Correlation test for fluxes averaged
over 1, 5 and 10 days, during the radio quiescent state \citepalias[see][Figure 1]
{cho02a}, also corresponding to X-ray quiescent and hard state. Though the value
of the SRC coefficient increases with increasing bin time, it should be noted that
the number of degrees of freedom  decreases and hence the best correlation result is
obtained for one day averaging (as seen by the value of the null hypothesis
probability). Hence the correlation time scale is shorter than a day. The X-ray
hardness ratio shows  a very strong and significant anti-correlation with the radio
emission. Interestingly this anti-correlation is stronger and more significant than
the anti-correlation of the hard X-ray flux with radio (or soft X-ray). This suggests
spectral bending being correlated with the radio emission, amply demonstrated in the
top panel of Figure \ref{fig3} \citepalias[see also Figure 3][]{cho02a}. The pivoting
of the spectrum around 12 keV, as obtained from pointed {\it RXTE} observations,
explains the anti-correlation between the soft (2 -- 10 keV, {\it RXTE}-ASM) and hard
X-ray (20 -- 100 keV {\it CGRO}-BATSE) emission.

We provide the X-ray wide band spectra following the same procedure as adapted in
\citetalias{cho02a} \citep[see also][]{vad01a}. The X-ray spectra of black hole
sources contain a thermal and a non-thermal part, which are conventionally modeled
as a disk black body spectrum and a power-law (or cut-off power-law) or more
realistic models incorporating Compton scattering from thermal as well as non-thermal
electrons \citep{zdz01}. Since our aim is to make a wide band description of the
spectra to understand the broad features, we have adopted an analytically simpler
model consisting of a disk black-body and Comptonization from thermal electrons
(\citealt{sun80} - the CompST model), plus an additional power law component required
to fit the spectra in the quiescent and hard state \citep{cho02b}. For this source
the X-ray spectra below 5 keV is dominated by the numerous photoionization emission
lines \citep{pae00} originating in the hot gas, probably from the dense stellar wind
of the Wolf-Rayet companion, engulfing the binary system  \citep{nak93}. Therefore
the continuum disk blackbody emission, contributing in the energy range below 5 keV
is obscured and hence needs more rigorous analysis and better resolved spectra, which
is beyond the scope of this paper. Therefore we present the spectra above 5 keV,
modeled with the CompST and power law component. \citet{vad02} have shown that
CompST model gives a functionally correct description of the more elaborate numerical
codes, although with slightly different parameters. Moreover, such composite models
have been used earlier for Cyg X-3  \citep{raj94}. Table \ref{tab2} gives the results
of the  spectral fitting, along with the soft X-ray (ASM), hard X-ray (BATSE) and
radio (2.2 \& 8.3 GHz, GBI) flux, of two observations corresponding to the extreme
behavior of the source within the precincts of the hard state.

In \citetalias{cho02a}, more spectral data were presented that showed a systematic
change in the wide band spectrum correlated with the radio flux. The available
spectra showed a pivoting behavior around 12 -- 15 keV. This fact, coupled with the
strong correlation between the soft X-ray and radio as well as that anti-correlation
of the hard X-ray flux with radio (and soft X-ray flux) strongly suggest that the
spectral shape governed by a pivoting behavior at around 12 keV is responsible for
the observed correlations.

\subsection{GRS 1915+105}
The X-ray binary GRS 1915+105 was first detected in 1992 \citep{cas92}, and has been
observed in the X-ray, radio and infra-red bands since then \citep[see][for a review]
{bel02}. \citet{mun01} classify the radio emission into three classes, radio faint,
radio steep and radio plateau, and find that radio emission is always present in the
hard steady X-ray state. This source is extremely variable in nature and it has been
classified into several variability classes \citep{bel00}. The $\chi$ class is the
closest analogue to the canonical low-hard states of Galactic black hole sources
\citep{rao00}. \citet{bel00} identify three stretches of long duration $\chi$ classes
and two of them ($\chi_2$ and $\chi_3$) have simultaneous BATSE and GBI observations.
These two periods are demarcated by numbers in the top panel of Figure \ref{fig1},
and are used for the correlation analysis. The results of the SRC test, given in
Table \ref{tab1} are similar to that of Cyg X-3. The radio and soft X-ray fluxes are
well correlated. The anti-correlation of  the hard X-ray flux with both radio and
soft X-ray flux is not as strong as in the case of Cyg X-3. The correlation time
scale, as can be concluded from the strength of the correlation, is one day or less.
The correlation between the X-ray hardness ratio and radio flux again gives results
similar to those of Cyg X-3, suggesting a spectral pivoting correlated to the radio
emission. Hence, it is evident that the radio X-ray correlation behavior in the
steady long term hard state of this source is similar to that of Cyg X-3.

Two unfolded spectra for GRS 1915+105 corresponding to GBI 8.3 GHz fluxes of 17 and
77 mJy respectively are overlaid in the bottom panel of Figure \ref{fig3}. Table
\ref{tab2} gives the details of the soft X-ray (ASM), hard X-ray (BATSE) and radio
(2.2 \& 8.3 GHz, GBI) flux along with the best fit parameters of the spectral fitting
\citep{vad01a}. It is interesting to note that the wide-band spectra at extreme radio
emissions shows a cross-over at higher energies (20 keV) compared to Cyg X-3 and, by
association, we suggest that a spectral pivoting occurring at higher energies is
responsible for the observed correlations. A possible reason for a weaker
anti-correlation between hard X-ray (20 -- 100 keV) flux and soft X-ray (2 -- 12 keV)
flux (and radio) compared to that in Cyg X-3 is that in the case of Cyg X-3 the soft
X-ray (ASM) and hard X-ray (BATSE) energy ranges are, correspondingly, below and
above the pivot energy of around 12 keV, whereas for GRS 1915+105, the pivot energy
is at higher energies of around 20 keV, and the spectrum is relatively harder.

\citet{rau03} have made a detailed study of all the $\chi$ state observations of GRS
1915+105 based on the analysis of four years of pointed {\it RXTE} PCA and HEXTE
observations. They find no correlation between radio and soft X-ray flux (1 -- 20
keV), however, they find an anti-correlation between radio and hard X-ray (20-100
keV) flux, and also find that the slope of hard power law spectrum correlates
positively with the radio flux in the low hard state of the source with observations
during high radio emission showing a softer spectrum. The latter results agree with
our findings presented here. \citet{rau03} also report spectral pivoting occurring
between 20 -- 30 keV in X-ray spectra in hard state of GRS 1915+105. The results of
\citet{rau03} are consistent with our findings, except that the radio and soft X-ray
fluxes are not correlated in their data, whereas we find a good correlation between
radio and soft X-ray flux. Our results are based on the ASM data which is not very
sensitive to X-rays above 10 keV, whereas \citet{rau03} use the flux up to 20 keV
using a model fit to the joint PCA and HEXTE observations. The lack of correlation
between radio:soft X-ray could be due to the spectral pivoting around 20 keV, because
of which the soft X-ray flux from i{\it RXTE} PCA data will comprise of both
correlated and anti-correlated fluxes thus weakening or averaging out the
correlation.

\subsection{Cyg X-1}
Cyg X-1 \citep{bow65} is the first Galactic black hole candidate \citep{her95} whose
optical counterpart, O9.7Iab super giant HDE 226868, was among the earliest to be
identified for an X-ray binary \citep{bol72, web72}. A persistent source in X-ray,
radio \citep{bra71} and optical, it shows a binary modulation with a period of 5.6
days in all the bands \citep{poo99, bro99}. The radio emission is weak, generally
around 15 mJy, varying between 10 to 25 mJy in the low-hard X-ray state getting
considerably weaker in the high-soft state of X-ray emission \citep{bro99}.

Figure \ref{fig2} gives the combined light curve of Cyg X-1 in the soft X-ray (2 --
12 keV, ASM, {\it top panel}), hard X-ray (40 -- 140 keV, BATSE, {\it middle panel})
and the radio (2.2 GHz, GBI, {\it bottom panel}). The region 1 as demarcated in the
top panel of the figure denotes the period of the long term low-hard state of the
X-ray emission. Since our emphasis is on the study of the long-term correlated radio:
X-ray behavior of this source in low-hard state, we consider only this period for the
SRC test. The results of the SRC test for Cyg X-1 are  given in Table \ref{tab1}. The
table shows that pattern of the correlation between X-ray and radio emission for Cyg
X-1 seems to be different than for Cyg X-3 and GRS 1915+105. The soft X-ray and radio
fluxes are not as well correlated as for the other two sources, specially for 1 day
averages, being only 0.29 compared to 0.68 and 0.56 for Cyg X-3 and GRS 1915+105
respectively. Nevertheless, the correlation is significant at the level of about one
part in 10$^6$. The anti-correlation of hard X-ray (40 -- 140 keV, BATSE) flux with
the soft X-ray (2 -- 10 keV) as well as radio (2.2 GHz, GBI) fluxes found for Cyg
X-3 and GRS 1915+105 is not present in Cyg X-1. Instead, the SRC test shows that the
hard X-ray positively correlates with the both soft X-ray and radio emission. Also
the X-ray hardness ratio does not show any correlation with radio flux in the case
of Cyg X-1, unlike for Cyg X-3 and GRS 1915+105.

\citet{bro99} report a value of the SRC coefficient for soft X-ray:radio flux
correlation in low-hard state of 0.3 for 1 day average, after removing the mean
orbital light curve, which is close to 0.29 found by us for the same correlation,
without removal of the orbital modulation effects. \citet{bro99} point out that
loose correlation of radio and soft X-ray fluxes (for their 1 day averages) may be
partly due to the possible offset between the radio and X-ray long period
($\sim$142 days) light curves. They also give the scatter plot of ASM, BATSE and
radio fluxes.

The strength of the SRC between the radio and the both soft and hard X-ray are
similar (Table \ref{tab1}), showing moderately strong correlation, whereas the soft
and hard X-ray flux show a very strong positive correlation, the SRC coefficient
being 0.70 for 1 day averages. Further, ASM and BATSE observations are at different
times during the day implying that intra-day variability is relatively weak compared
to variability on longer time scales as both fluxes are strongly correlated over
longer time scales. Clearly, the similarity between hard X-ray:radio and soft X-ray:
radio flux correlation is because of strong correlation between hard and soft X-ray
fluxes. As mentioned earlier, the radio  emission in Cyg X-1 is quite weak (around
15 mJy) varying between 10 and 25 mJy in the low-hard state. As the  GBI
observations have an error of nearly 4 mJy, a detailed wide-band spectral analysis
for two extreme values of radio flux, for finding the relation between the shape of
the X-ray spectra and the radio emission and the pivoting behavior and the pivot
energy, as done for Cyg X-3 and GRS 1915+105, is difficult to be carried out for
Cyg X-1. 

\citet{zdz02} have shown that the long term variability of the X-ray emission from
this source in hard state comprises of two types of spectral variability, one
corresponds to the change in the shape of the spectra (with spectral shape pivoting
around 80 keV) with change in soft X-ray flux and the other corresponds to the
change in total flux, with the spectra simply moving up and down parallel to each
other with a constant shape, in the whole X-ray broad band. This may explain the
lack of correlation between the X-ray hardness ratio and the radio flux as well as
the comparative weakness of the strength of the SRC correlation between ASM and
radio flux. \citet{zdz02} have analyzed the various correlations among the fluxes of
the three energy channels of ASM (1.5 -- 3, 3 -- 5 and 5 --12 keV) along with the 20
-- 100 keV and 100 -- 300 keV flux of BATSE and the corresponding specific spectral
index of these bands. They conclusively show that in the low-hard state of the X-ray
emission, over long periods, the change in the spectral shape occurs with a pivoting
around 50 -- 90 keV. This explains that the BATSE flux, being dominated by the lower
energy photons is very strongly correlated to the ASM flux. The lack of
anti-correlation between the X-ray flux hardness ratio and the radio emission may
also be explained by this fact, as the lower energy of the BATSE flux, below the
pivot point, is correlated to the radio emission along with the ASM flux. 


\section{Discussion}

\subsection{Comparison with earlier results }
The results presented in this paper will be discussed along with similar results for
GX 339-4 reported by \citet{cor00,cor03}. The compact object in GX 339-4 is also
believed to be a black hole \citep{hyn03}. \citet{cor00} find that for GX 339-4, the
radio flux is strongly correlated with both soft and hard X-ray, covering the range
3-200 keV, in low-hard state, similar to the results obtained by us for Cyg X-1.

If we consider the results of the radio, soft X-ray and hard X-ray correlation
analysis for these four sources, at first glance no consistent picture of the
correlated variability pattern emerges. However, it is immediately noticeable that
Cyg X-3 and GRS 1915+105 have similar overall correlation pattern. For both, the soft
X-ray flux is correlated with the radio flux, and the hard X-ray flux is
anti-correlated with the both radio and soft X-ray flux. Additionally, the hardness
ratio is also strongly anti-correlated with the radio flux. Wide band X-ray spectral
analysis in the hard state for both the sources at different radio flux levels
suggests  pivoting of the spectrum around 10 -- 25 keV correlated with the radio
emission.

It can also be noticed that correlation among the radio, soft and hard X-ray fluxes
for Cyg X-1 and GX 339-4 is similar. Both show a positive correlation among the
radio, soft X-ray and hard X-ray fluxes. For Cyg X-1 the hardness ratio is not
correlated with radio flux. Further, \citet{zdz02} find a pivoting of the X-ray
spectrum  of Cyg X-1 at higher energy of around 50-90 keV. Similarly, the wide band
X-ray to gamma-ray spectral analysis of GX 339-4 \citep{war02} has shown that there
is a pivoting in the spectrum at energies $\sim$300 keV in the low-hard state of the
source.
   
At this stage it will be worthwhile to note other similarities in the X-ray and
radio emission characteristics of Cyg X-3 and GRS 1915+105 {\it vis a vis} Cyg X-1
and GX 339-4. Most notable are that the first two sources are the strongest and most
variable radio sources amongst the Galactic X-ray binaries whereas both Cyg X-1 and
GX 339-4 are amongst the comparatively weak and steady radio sources. On the other
hand, both Cyg X-1 and GX 339-4 have a very hard X-ray spectrum compared to Cyg X-3
and GRS 1915+105.

Thus both X-ray sources with softer X-ray spectrum  have a lower pivot energy and
Cyg X-1 with a much harder X-ray spectrum has a much higher pivot energy, indicating
that the pivot energy is directly related to spectral shape. The correlation between
hard X-ray flux and radio and soft X-ray fluxes observed in Cyg X-1 and GX 339-4 can
then be explained because hard X-ray flux (40 -- 140 keV for Cyg X-1, 20-200 keV for
GX 339-4) is around or below the pivot energy in these two sources and will,
therefore, be correlated with the soft X-ray and thereby the radio fluxes. Thus, it
is quite evident that X-ray fluxes below and above the pivot energy are
anti-correlated for these X-ray sources and the reported differences between radio,
soft X-ray and hard X-ray correlation amongst these X-ray sources is an instrumental
artifact where the {\it RXTE}-ASM and {\it CGRO}-BATSE energy ranges are fixed and
the pivot energy varies from source to source. This is supported by finding of
\citet{zdz02} who report an anti-correlation between 1.5 -- 3.0 keV and 100 -- 300
keV flux and find a very weak correlation between 1.5 -- 3.0 keV and 20 -- 100 keV
flux for Cyg X-1 in the low-hard state. Thus the X-ray -- radio behavior of these
X-ray sources are consistent in terms of correlation between soft X-ray, hard X-ray
and radio emissions reported here.

\subsection{The X-ray soft state and suppressed radio emission}
In the previous sub-section we have shown that the X-ray radio emission
characteristics in the hard states of the highly variable sources Cyg X-3 and GRS
1915+105 are similar to those seen in the low-hard states of the well studied black
hole sources Cyg X-1 and GX 339-4, once we assume different pivot energy correlated
to the radio emission. Here we explore whether the suppressed radio emission seen in
the high-soft states of Cyg X-1 \citep{bro99} and GX 339-4 \citep{cor00} are seen in
these two sources. We must caution that the identification of various spectral states
using monitoring data is fraught with difficulties of flaring emissions which could
be quite delayed in the various emission bands.

\begin{figure}
\figurenum{4}
\epsscale{0.5}
\plotone{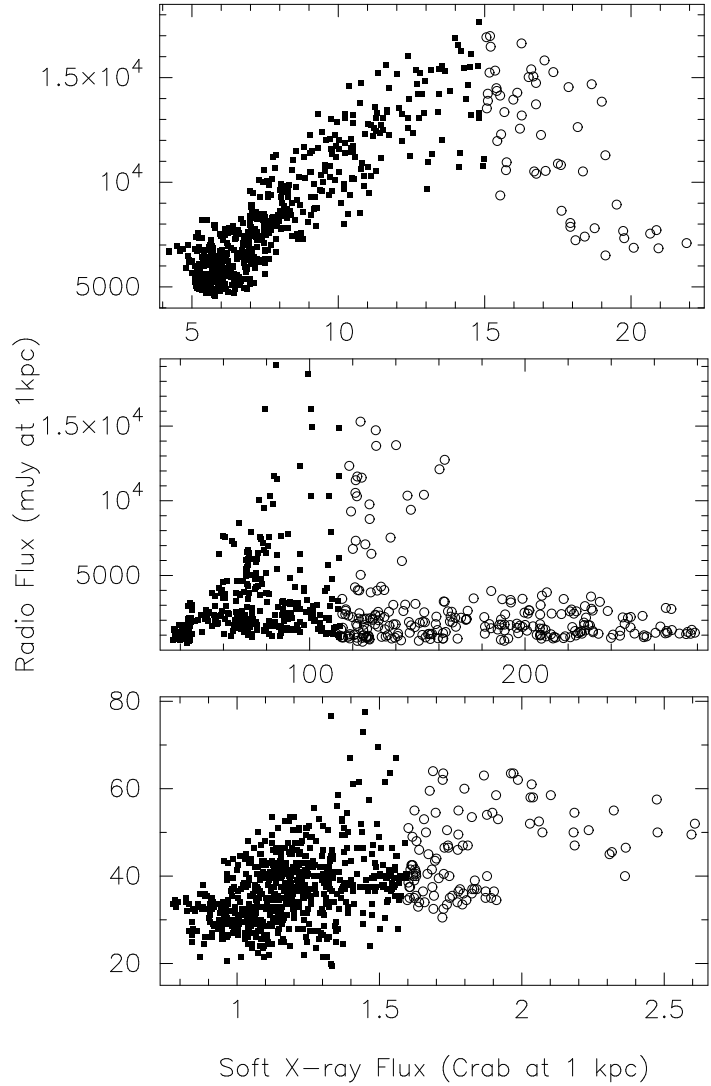}
\caption{The radio (GBI, 2.2GHz) and soft X-ray ({\it RXTE}-ASM, 2-12 keV) emission scatter diagram of Cyg X-3 ({\it top panel}), GRS 1915+105 ({\it middle panel}) and Cyg X-1 ({\it bottom panel}) for the long-term, steady, hard ({\it filled squares}) as well as soft ({\it open circles}) states, after removing the data for the flaring states. \label{fig4}}
\end{figure}

\citet{cor00} show that with the X-ray state transition from low-hard to high-soft
the radio emission evolves from a jet like synchrotron emission to quenched emission
in GX 339-4. This state is generally preceded by a low-hard X-ray (with correlated
radio emission) and followed by an X-ray off state (with radio off). For a
comparative analysis we plot the radio (GBI, 2.2GHz) and soft X-ray ({\it RXTE}-ASM,
2-12 keV) scatter diagram in Figure \ref{fig4} for Cyg X-3 ({\it top panel}), GRS
1915+105 ({\it middle panel}) \& Cyg X-1 ({\it bottom panel}), for the non flaring
states, which include hard as well as soft states. We have attempted to distinguish
the high and low states by the soft X-ray flux and denoted them by open and filled
symbols, respectively. For Cyg X-3 the major flares are excluded, and in the process
we have excluded the (very low) quenched radio emission immediately preceding the
major flares. For GRS 1915+105, too, the data for the radio flares are excluded. It
is evident that even for these two sources the radio emission is suppressed in the
high state, analogous to the canonical high-soft state. Cyg X-3 shows a very
systematic behaviour, with the radio positively correlated to the soft X-ray in the
hard state, until it transits to the soft state, where the radio emission is
negatively correlated to the soft X-ray emission. For GRS 1915+105 the transition
into the soft state with suppressed radio emission is not that drastic but definitely
pronounced. Cyg X-1 shows a more scattered association between the radio and X-ray
emission, but the suppression of the radio emission with higher ASM flux is evident.
Hence, it can be comfortably claimed that the suppression of the radio emission with
the X-ray state transition is a generally consistent feature of the X-ray binary
systems (BHCs), irrespective of their individual spectral characteristics. Therefore
a consistent picture of the accretion-ejection picture is emerging from the
observational analysis of sources with apparently very diverse behavioural patterns.

\subsection{A universal correlation and its origin}
In the previous two sub-sections we have shown that the four sources, viz. Cyg X-3,
GRS 1915+105, Cyg X-1 and GX 339-4, all show a consistent picture of
accretion-ejection mechanism, with the radio emission correlated to the X-ray
spectral pivoting in the low state and suppression of the radio emission in the high
state following the X-ray state transition. The most notable feature, however, is the
strong positive correlation between the radio and soft X-ray flux in the hard state
of all the sources. This is similar to the correlation reported for GX 339-4
\citep{han98, cor00, gal02} and V404 Cyg \citep{gal02}). We explore below whether the
observed correlation is an universal phenomena among the Galactic black hole
candidate sources.

\begin{figure}
\figurenum{5}
\epsscale{0.5}
\plotone{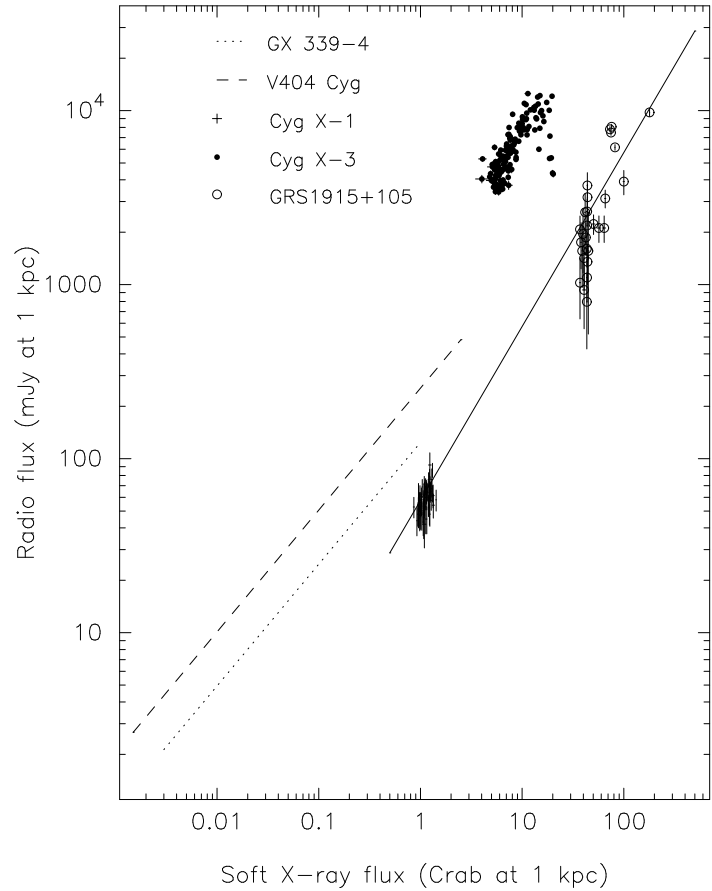}
\caption{A plot of radio flux at 2.2 GHz (based on GBI data) normalized to a
distance of 1 kpc against the soft X-ray flux in 2 -- 12 keV (based on {\it RXTE}-ASM
data) normalized to Crab at 1 kpc for Cyg X-1, Cyg X-3 and GRS 1915+105, all in their
corresponding low-hard states. The data are averaged for 5 days. The power-law fit
(with an index of 0.7) reported for GX 339-4 and V404 Cyg by Gallo et al. (2002) are
shown as dotted and dashed lines, respectively. The continuous line is a linear fit
to the combined data of Cyg X-1 and GRS 1915+105. \label{fig5}}
\end{figure}

In Figure \ref{fig5}, we show a scatter plot of the radio flux against the soft X-ray
flux for Cyg X-1 (plus sign), Cyg X-3 (filled circles) and GRS 1915+105 (open
circles), all in their corresponding low-hard and their analogous states. The data
are normalized to a distance of 1 kpc, with the assumed distances of 2 kpc, 8.5 kpc
\& 12.5 kpc, respectively for the above three sources. For the soft X-ray flux (based
on {\it RXTE}-ASM data)  75 ASM counts s$^{-1}$ is taken as the observed Crab flux.
Individual data points are the average value for a bin size of 5 days. Recently,
\citet{gal02} have detected a correlation between the radio flux (S$_{radio}$) and
X-ray flux (S$_X$) of the form  S$_{radio}$ = k S$_X^{+0.7}$ for GX 339-4 and V404
Cyg, all the way from the quiescent level to close to high-soft state transition.
This relationship is also shown in Figure \ref{fig5} as a dotted line (for GX 339-4)
and dashed line (for V404 Cyg). The extents of the lines correspond to the data used
for the fit in \citet{gal02}.

The remarkable feature of Figure \ref{fig5} is that a simple relation seems to hold
for all the black hole sources over close to 5 orders of magnitude variation in the
luminosity, in the low-hard state. The data points for Cyg X-1 fall just below the
correlation found for GX 339-4, whereas, the data points for GRS 1915+105 lie on the
line extrapolated from the correlation derived for V404 Cyg. The data points for
Cyg X-3 are parallel to those of GRS 1915+105. It should be noted here that Cyg X-3
shows a very strong orbital modulation which is most probably due to the obscuration
of the accretion disk by a cocoon of matter surrounding the accretion disk. Hence
the `true' X-ray emission from the accretion disk should be larger than the observed
one and the data points should move closer to the extrapolated line from GX 339-4
and V404 Cyg. Since Cyg X-3 is suspected to be a micro-blazar \citep{kul01}, it is
also quite possible that the radio emission is over-estimated due to strong beaming
and Doppler boosting. It is also noteworthy that GRS 1915+105, the most massive
stellar mass black hole known, has the highest intrinsic X-ray emission.

Fitting individual data points of the sources with a function of the form
S$_{radio}$ = k S$_X^{+0.7}$ gives the value of the constant term as 54 mJy, 235 mJy
and 1376 mJy, respectively for Cyg X-1, GRS 1915+105 and Cyg X-3. These values should
be compared to those obtained for GX 339-4 and V404 Cyg, 124 mJy and 254 mJy,
respectively. The individual data points, however, are also consistent with a linear
relation and the continuous line in the Figure \ref{fig5} is a linear fit to the
combined data of Cyg X-1 and GRS 1915+105.

Although we cannot completely rule out the possibility that the individual
correlations of these sources may exist due to processes unrelated to one another,
the fact that the data points for sources close to their `off' states (GX 339-4 and
V404 Cyg) occupy one extreme of observations while sources with repeated radio flares
(Cyg X-3 and GRS 1915+105) occupy the other extreme strongly suggests of a common
physical mechanism in operation in all the sources. One natural consequence to be
explored is whether the radio and the X-ray emissions are emitted directly from the
same source region. Since there are very strong observational evidences for the
radio emission to be of synchrotron origin, it is tempting to assume that the X-ray
emission too is emitted by the same process, but at the base of the jet. There are
evidences for X-ray synchrotron emission being responsible for the X-ray spectrum of
some black hole sources like XTE J1118+480 \citep{mar01} and part of the spectrum
in sources like GRS 1915+105 \citep{vad01a}. However, for several black hole sources
wide-band X-ray spectrum has been extensively studied and the spectral shape is
inconsistent with a simple synchrotron emission. Further, the X-ray:radio correlation
appears to be valid from `off' state to low-hard and the associated analogous states,
and all the way up to the intermediate state, and it is extremely unlikely that the
bulk of the X-ray emission is due to synchrotron emission in all these states. 

Recently \citet{mar03} have suggested jet synchrotron emission as a possible way to
explain the broadband (including X-ray) features of GX 339-4. To explain the observed
correlation, their model predicts that the X-ray emission is mostly due to
synchrotron emission (with a power-law spectral shape). Since the wide band hard
X-ray/ low energy gamma-ray spectra of black hole sources in their hard states like
GX 339-4 \citep{war02}, GRS 1915+105 \citep{zdz01}, and Cyg X-1 \citep{zdz00} require
thermal-Compton emission to explain the spectral shape, we explore below alternate
models where the X-ray emission is primarily due to accretion disk emission.

\subsection{X-ray spectral shape as the ``driver'' of the radio emission}
The wide band X-ray spectral shapes of Galactic X-ray binaries with black holes as
compact objects show a systematic and predictable behavior, particularly in well
studied sources like Cyg X-1 \citep{zdz00} and GX 339-4 \citep{war02}, where,
in the low-hard state, the spectral energy distribution peaks at $\sim$ 100--300 keV,
with the emission being dominated by thermal-Compton emission from a population of
hot electrons. For sources which go to the `off' state the X-ray spectrum is quite
similar to the low-hard state, but at a very low intensity. In the intermediate state
the soft X-ray flux increases and in the high-soft state the spectrum is dominated by
thermal emission from the accretion disk. Although it is quite convenient to assume
that the mass  accretion rate is responsible for such systematic changes, there is no
strong  evidence for this \citep[see][for a different behaviour of XTE J1550-654]
{hom01}. Nevertheless, one can safely conclude that some unspecified `accretion
parameters' causally affect the wide band X-ray spectral shape of black hole X-ray
binaries.

The evidences presented in this paper strongly suggests that the very same `accretion
parameters' must be causally responsible for the radio emission, rather than the
amount of soft X-ray emission, provided they account for the suppressed radio
emission in the soft state. Such a hypothesis neatly explains the behavior of Cyg
X-3, which is quite bright in all the three energy ranges and hence the observational
uncertainty is quite low. It can also be noticed from Table 1 that the most
significant correlation for Cyg X-3 is between the radio emission and the ratio of
hard X-ray flux to soft X-rays. Though such an explanation is not very clear in other
sources, all the available observations are consistent with this. Since the radio
emission too shows increasing emission from `off' state to low-hard state (in GX
339-4 and V404 Cyg), correlated behavior in low-hard state (the above two sources and
Cyg X-1), and high radio emission in an intermediate state very close to the high
state (in GRS 1915+105 and Cyg X-3) and suppressed radio emission in the high state
(Cyg X-3, GRS 1915+105, Cyg X-1 \& GX 339-4), we speculate that the soft X-ray
intensity determines the spectral shape and the accretion disc condition in these
sources, which in turn determines the amount of radio emission (in the X-ray
quiescent hard state of these sources). The transition of the systems (Cyg X-3 \&
GRS 1915+105) into flaring state and their corresponding behavior in the radio as
well as X-ray bands is an issue not discussed here. 

\subsection{The TCAF model for the X-ray:radio correlation}

Though there are models describing the  accretion disk emission \citep{zdz00} or jet
emission \citep{mar03}, there are very few models which self-consistently solve the
accretion and ejection phenomena seen in black hole sources. Since our findings
suggest a close connection between these two phenomena, we attempt below to
qualitatively explain the X-ray radio association using one such model: the Two
Component Advective Flow (TCAF) model of \citet{cha96}. According to this model, the
Compton scattered X-rays in a black hole source originates from a region close to the
compact object, confined within the Centrifugal Boundary Layer (CENBOL). The X-ray
spectral shape in various `states' of the source essentially depends on the location
of the CENBOL and a detailed description can be found in \citet{cha95} and
\citet{ebi96}. At low accretion rates, the CENBOL is far away from the compact
object and the X-ray spectrum is dominated by a thermal-Compton spectrum, originating
from the high temperature region within the CENBOL. In the transition state, the
CENBOL comes closer to the compact object and the CENBOL can sometimes give rise to
radial shocks, causing intense quasi-periodic oscillations, as seen in GRS 1915+105.
In the high state, the increased accretion rate produces copious photons in the
accretion disc which cool the Compton region, giving rise to very intense disk
blackbody emission along with bulk motion Comptonization (a power-law in hard X-rays
with a photon index of $\sim$2.5). At some critical accretion rates, the state
transitions could be oscillatory as seen in GRS 1915+105 \citep{cha00}.

The behavior of TCAF disks and the outflow has been studied in detail in
\citet{das98}. The outflow rate is found to be a monotonic function of the
compression ratio, R,  of the gas at the shock region. In this scenario, at low
accretion rates, the CENBOL is far away from the compact object, and a weak shock can
form with low compression ratio, giving low and steady outflow. If this outflow
gives rise to radio emission, one can expect a relation between the radio emission
and the X-ray emission. In this state (off state to low-hard state), an increased
accretion rate increases the overall amount of energy available to the Comptonizing
region and hence increasing the X-ray emission. The CENBOL location would be pushed
inward, increasing the compression ratio (and hence increasing the radio emission)
and also can increase the temperature and optical depth of the Comptonizing region,
thus giving rise to a pivoting behavior at hard X-rays (50 -- 90 keV) as seen in
Cyg X-1 and GX 339-4. At increased accretion rate, the CENBOL can come closer to
the compact region, giving the spectral and radio properties as seen in GRS 1915+105
and Cyg X-3. For a given accretion rate the compression ratio, after reaching a
critical value (with the shock region coming correspondingly closer to the event
horizon), causes the source to transit into the high-soft state state, for which the
radio emission is progressively suppressed \citep{cha99}. This model qualitatively
explains all the observed X-ray spectral and radio properties of Galactic black hole
sources presented here.  

\section{Summary and conclusions}
A complete understanding of the accretion-ejection physics in Galactic microquasars
demands proper interpretation and modeling of all the varied states of X-ray and
radio emission, encompassing the various flaring and steady emissions covering all
ranges of time scales. In this paper we have taken a first step in this direction by
attempting to understand the long term variation of the (non flaring) radio emission
associated with the X-ray emission in the steady hard and soft states. We have
analyzed the (quasi) simultaneous observations on GRS 1915+105 and Cyg X-1 using the
{\it RXTE}-ASM, {\it CGRO}-BATSE and GBI data and made a detailed study of
correlation between radio and X-ray fluxes. Based on this analysis along with
discussion of earlier published results on Galactic microquasars we find that:

\begin{itemize}
\item
A correlation exists between the soft X-ray and radio emission of GRS 1915+105 based
on the data during the long $\chi$ state (associated to the low-hard  state). The
hard X-ray emission is anti-correlated with both radio and soft X-rays. There is a
spectral pivoting at around 20 keV, correlated with the radio flux. 
\item
Comparing these results with those of \citet{rau03} who found a strong correlation
between radio emission and the X-ray spectral index in the $\chi$ states, we conclude
that the X-ray and radio emission characteristics of GRS 1915+105 are similar to
those of Cyg X-3 \citepalias{cho02a}. The only difference lies in the values of the
pivot energy of the X-ray spectra, which is around $\sim$12 keV in Cyg X-3 and around
$\sim$20 keV in GRS 1915+105.
\item
A three way correlation among soft X-ray, hard X-ray and radio emission has been
found in the low-hard state of Cyg X-1, confirming the results of \citet{bro99}.
Comparing this result with those of \citet{zdz02} who have found that soft X-ray and
hard X-ray above 100 keV are anti-correlated and also that there is a spectral
pivoting at around 50 -- 90 keV, we conclude that the X-ray:radio behavior of Cyg X-1
is similar to that of Cyg X-3 and GRS 1915+105, but for the fact that the pivoting
energy is at a higher value.
\item
The X-ray:radio properties of Cyg X-1 are quite similar to that of GX 339-4, where a
3-way correlation between soft X-ray, hard X-ray and radio emission has been reported
\citep{cor00,cor03}. Though an anti-correlation between soft X-ray/radio with hard
X-rays has not been reported in this source, we note that the X-ray spectrum during
low-hard state too shows a pivoting behavior at high energies $\sim$300 keV \citep{war02}.
\item
The radio emission is suppressed for Cyg X-1 \& GX 339-4 in their high-soft state
and similarly for Cyg X-3 and GRS 1915+105 in their high states (with associated
softer spectra). Therefore, all these four sources with apparent diverse X-ray and
radio properties show very similar behavioural pattern encompassing the long term
steady non-flaring state.
\item
Compiling the soft X-ray and radio observations of the above sources (GRS 1915+105,
Cyg X-3, and Cyg X-1) with the published correlation in GX 339-4 and V404 Cyg
\citep{gal02}, we find that all the sources show a monotonic increase of radio
emission with the soft X-ray emission, spanning a 5 orders of magnitude variation in
their intrinsic luminosities. Cyg X-3 deviates from a single relation by about an
order of magnitude which can be reconciled if 1) the observed X-ray intensity is an
under-estimate because of obscuration and/or 2) the observed radio intensity is an
over-estimate because of beaming and Doppler boosting.
\item
If a common physical phenomena is responsible for such an uniform relation spanning
across `off' state to intermediate state, we argue that both radiations (X-ray and
radio) are unlikely to be originating from a single mechanism like synchrotron
emission.
\item
Finally, we invoke a  Two Component Advective Flow (TCAF) model \citep{cha96} to
explain the accretion-ejection behaviour in these systems in the steady hard as well
as soft states.
\end{itemize}

\section*{Acknowledgements} This research has made use of data obtained through the 
HEASARC Online Service, provided by the NASA/GSFC, and the Green Bank Interferometer,
a facility of the National Science Foundation operated by the NRAO in support of NASA
High Energy Astrophysics Programs. The authors thank the anonymous referee for the
extensive and insightful comments. MC and SVV have been partially supported by the
Kanwal Rekhi Scholarship for Career Development. AKJ is grateful to P. S. Goel,
Director, ISAC and K. Kasturirangan, Chairman, ISRO, for their constant encouragement
and support during the course of this work.

\clearpage

\begin{deluxetable}{lccccccccc}
\tablecolumns{10}
\tablewidth{0pc}
\tablecaption{The Spearman Rank Correlation (SRC) coefficient, null-hypothesis 
probability and D-parameter among 1) the radio, soft X-ray and hard X-ray fluxes
and 2) the hardness ratio of X-ray (ratio of BATSE to ASM flux), the radio and the
soft X-ray fluxes, for Cyg X-3, GRS 1915+105 and Cyg X-1 in the low-hard
state with the observed fluxes averaged for 1, 5 and 10 days. \label{tab1}}
\tablehead{
\colhead{} & \multicolumn{3}{c}{\bf Cyg X-3} & \multicolumn{3}{c}{\bf GRS1915+105} & \multicolumn{3}{c}{\bf Cyg X-1}\\
\cline{1-10} 
\colhead{} & \colhead{SRC} &  \colhead{Null Prob.} & \colhead{D-Par.} & \colhead{SRC} & \colhead{Null Prob.} & \colhead{D-Par.} & \colhead{SRC} & \colhead{Null Prob.} & \colhead{D-Par.}
}
\startdata
\multicolumn{1}{l}{1 day avg. flux} &  \multicolumn{3}{c}{no. of data points: 532} & \multicolumn{3}{c}{no. of data points: 108} & \multicolumn{3}{c}{no. of data points: 268} \\
\cline{1-1} \cline{3-3} \cline{6-6} \cline{9-9}
ASM:GBI & 0.68  & 0 & 15.9 & 0.61  & 3.6$\times10^{-12}$ & 6.8 & 0.29 & 1.6$\times10^{-6}$ & 1.4\\
GBI:BATSE & $-$0.43 & 7.2$\times10^{-26}$ & $-$3.7 & $-$0.27 & 4.1$\times10^{-3}$ & $-$1.8 & 0.33 & 1.6$\times10^{-1}$ & 1.4 \\
ASM:BATSE & $-$0.48 & 1.9$\times10^{-32}$ & $-$6.7 & $-$0.23 & 1.8$\times10^{-2}$ & $-$0.8 & 0.70 & 2.4$\times10^{-40}$ & 13.1 \\
RATIO:GBI & $-$0.65 & 0 & $-$3.9 & $-$0.62 & 4.0$\times10^{-13}$ & $-$4.1 & 0.01 & 9.3$\times10^{-1}$ & 2.5 \\
\multicolumn{1}{l}{5 day avg. flux} & \multicolumn{3}{c}{no. of data points: 149} & \multicolumn{3}{c}{no. of data points: 32} & \multicolumn{3}{c}{no. of data points: 65}\\
\cline{1-1} \cline{3-3} \cline{6-6} \cline{9-9}
ASM:GBI & 0.76 & 9.9$\times10^{-29}$ & 8.1 & 0.71 & 6.6$\times10^{-6}$ & 4.4 & 0.53 & 6.6$\times10^{-6}$ & 2.0 \\
GBI:BATSE & $-$0.61 & 1.3$\times10^{-16}$ & $-$2.5 & $-$0.33 & 6.7$\times10^{-2}$ & $-$1.3 & 0.53 & 6.5$\times10^{-6}$ & 2.0 \\
ASM:BATSE & $-$0.68 & 2.5$\times10^{-21}$ & $-$5.3 & $-$0.23 & 2.2$\times10^{-1}$ & $-$0.1 & 0.72 & 1.5$\times10^{-11}$ & 5.6 \\
RATIO:GBI & $-$0.71 & 3.9$\times10^{-31}$ & $-$2.0 & $-$0.65 & 5.0$\times10^{-5}$ & $-$1.5 & 0.03 & 8.4$\times10^{-1}$ & 1.6 \\
\multicolumn{1}{l}{10 days avg. flux} & \multicolumn{3}{c}{no. of data points: 75} & \multicolumn{3}{c}{no. of data points: 19} & \multicolumn{3}{c}{no. of data points: 33}\\
\cline{1-1} \cline{3-3} \cline{6-6} \cline{9-9}
ASM:GBI & 0.83 & 2.1$\times10^{-20}$ & 6.2 & 0.67 & 1.8$\times10^{-3}$ & 3.1 & 0.58 & 4.0$\times10^{-4}$ & 1.4 \\
GBI:BATSE & $-$0.72 & 4.7$\times10^{-13}$ & $-$1.5 & $-$0.23 & 3.2$\times10^{-1}$ & $-$0.8 & 0.61 & 1.0$\times10^{-4}$ & 2.0 \\
ASM:BATSE & $-$0.79 & 4.1$\times10^{-17}$ & $-$3.8 & $-$0.13 & 6.0$\times10^{-1}$ & 0.2 & 0.72 & 2.8$\times10^{-6}$ & 3.4 \\
RATIO:GBI & $-$0.85 & 1.9$\times10^{-22}$ & $-$2.1 & $-$0.58 & 9.2$\times10^{-3}$ & $-$0.5 & 0.11 & 5.5$\times10^{-1}$ & 1.4 \\
\enddata
\end{deluxetable}

\begin{deluxetable}{lcc}
\tablecolumns{3}
\tablewidth{0pc}
\tablecaption{The observed fluxes and X-ray spectral parameters of Cyg X-3 and GRS 1915+105 during two pointed {\it RXTE} observations, corresponding to the extreme
behaviour of the sources within the precincts of the respective low-hard states.
\label{tab2}}
\tablehead{
\colhead{} & \multicolumn{2}{c}{{\bf Cyg X-3}}
}
\startdata
MJD & 50717 & 50954\\
\cline{1-3}
Flux &  &  \\
\cline{1-1}
ASM (cts s$^{-1}$) & 11.11  & 5.37\\
BATSE (ph cm$^{-2} s^{-1}$) & 0.038 & 0.058\\ 
GBI-2.2 GHz (mJy) & 115 & 43\\
GBI-8.3GHz (mJy) & 165 & 53\\
\cline{1-3}
\colhead{Best fit parameters} &\multicolumn {2}{c}{CompST+powerlaw} \\
\cline{1-3}
kT$_e$ (keV) & 5.09$\pm0.38$ & 4.87$\pm0.08$\\
$\Gamma_X$ & 2.55$\pm0.22$ & 2.01$\pm0.04$\\
$\chi^2_\nu$(d.o.f.) & 0.74(86) & 1.42(108)\\
\cline{1-3}
\colhead{} &\multicolumn{2}{c}{{\bf GRS 1915+105}}\\
\cline{1-3}
MJD & 50421 & 50737\\
\cline{1-3}
Flux & & \\
\cline{1-1}
ASM (cts s$^{-1}$) & 38.59  & 34.96\\
BATSE (ph cm$^{-2} s^{-1}$) & 0.140 & 0.068\\ 
GBI-2.2 GHz (mJy)\tablenotemark{1} & 29 & 42\\
GBI-8.3 GHz (mJy)\tablenotemark{1} & 17 & 77\\
\cline{1-3}
\colhead{1}{l}{Best fit parameters} &\multicolumn {2}{c}{diskBB+CompST(+powerlaw\tablenotemark{2})} \\
\cline{1-3}
kT$_e$ (keV) & 20.76$^{+1.11}_{-0.97}$ & 4.89$^{+0.09}_{-0.08}$\\
kT$_in$(keV) & 1.28$^{+0.29}_{-0.36}$ & 1.99$^{+0.05}_{-0.04}$\\
$\Gamma_X$ & -- & 2.49$\pm0.01$\\
$\chi^2_\nu$(d.o.f.) & 1.14(121) & 1.55(121)\\
\enddata
\tablenotetext{1}{There is no observation on MJD 50421 for GRS 1915+105, we are quoting values for MJD 50422}
\tablenotetext{2}{The powerlaw component is present for MJD 50737, the radio loud $\chi$ state}
\end{deluxetable}

\clearpage

\end{document}